\documentclass[aps,prl,reprint,groupedaddress,superscriptaddress]{revtex4-1}
\usepackage[dvipdfmx]{graphicx,color}

\begin{document}


\title{Relaxation Dynamics of a Granular Pile on a Vertically-Vibrating Plate}

\author{Daisuke Tsuji}
\email{tsuji.daisuke@d.mbox.nagoya-u.ac.jp}
\affiliation{Department of Earth and Environmental Sciences, Nagoya University, Furocho, Chikusa, Nagoya 464-8601, Japan}

\author{Michio Otsuki}
\affiliation{Department of Materials Science, Shimane University, Matsue 690-8504, Japan}

\author{Hiroaki Katsuragi}
\affiliation{Department of Earth and Environmental Sciences, Nagoya University, Furocho, Chikusa, Nagoya 464-8601, Japan}

\date{\today}

\begin{abstract}
Nonlinear relaxation dynamics of a vertically-vibrated granular pile is experimentally studied.
In the experiment, the flux and slope on the relaxing pile are measured by using a high-speed laser profiler.
The relation of these quantities can be modeled by the nonlinear transport law \textcolor{red}{assuming the uniform vibro-fluidization of an entire pile.}
The fitting parameter in this model is only the relaxation efficiency, which characterizes the energy conversion rate from vertical vibration into horizontal transport.
We demonstrate that this value is a constant independent of experimental conditions.
The actual relaxation is successfully reproduced \textcolor{red}{by the continuity equation with the proposed model.
Finally its specific applicability toward an astrophysical phenomenon is shown.}
\end{abstract}


\maketitle

\begin{figure}[b]
\begin{center}
\includegraphics[width=80mm]{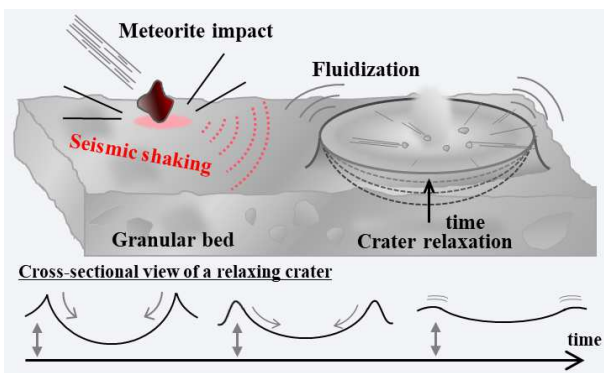}
\end{center}
\caption{\textcolor{red}{Schematic illustration of crater relaxation caused by meteorite-impact-induced seismic shaking on small astronomical objects.}}
\label{fig:application}
\end{figure}

A granular pile is usually stable when its slope is less than the angle of repose $\theta_{\rm c}$.
However, the structure of a pile can be relaxed by adding perturbations such as vibration.
In this sense, granular-heap structure is metastable.
The truly stable state under gravitational field must have a horizontally flat surface.
Therefore, when subjected to perturbations, a pile shows relaxation toward horizontal structure.
This type of granular relaxation is ubiquitous in many natural phenomena.
\textcolor{red}{For instance, the terrain development on astronomical objects covered with granular beds called regolith (e.g., asteroid Itokawa) is governed by the granular-heap relaxation~\cite{Michel2009}.
As illustrated in Fig.~\ref{fig:application}, meteorite-impact-induced seismicity is considered as a main perturbation source, which could erase characteristic terrains such as craters~\cite{Richardson2004}.}

In general, relaxation itself has long been an important process in complex soft matter physics~\cite{Ngai2011}.
One of the simplest modeling for relaxation is Fickian (liner) diffusion, which cannot explain the granular-heap relaxation~\cite{Roering2001,Sanchez2007}.
While several models have been proposed so far~\cite{Roering1999,Roering2001,Sanchez2007}, the parameter dependence of the relaxation process has not been investigated systematically.
Obviously, the relaxation property depends on various experimental conditions such as perturbation strength.
Revealing the parameter dependence is necessary for both the fundamental understanding of the relaxation process and its application to natural phenomena.
In this study, we perform systematic experiments and build a model which can quantitatively describe the granular-heap relaxation including the experimental parameter dependence.
\textcolor{red}{Besides, this paper briefly addresses the relaxation of craters by seismic shaking on a small asteroid.}

So far, steady granular flows on a pile with angle steeper than $\theta_{\rm c}$ have mainly been studied~\cite{GDR2004,Jop2005,Jop2006,Katsuragi2010}.
Instead, this study directly investigates the (non-steady) relaxation of a vibro-fluidized granular pile.
Using a high-speed laser profiler, spatiotemporally-resolved surface profiles of a relaxing pile are precisely obtained.
Based on the experimental data, a granular transport law called the Nonlinear Diffusion Transport (NDT) model is derived.
We show that the NDT model can explain the entire relaxation process of the pile with only one universal parameter representing the relaxation efficiency.

\begin{table}[b]
\setlength{\tabcolsep}{2mm}
\begin{center}
\begin{tabular}{lcccc}
\hline
Material   	  &  $d$~(mm)     &	$\rho$~(g/cm$^3$)	& tan$\theta_{\rm c}$ 	& Note	 \\
\hline
1. Almina ball    &	$0.5\pm0.1$ &	3.9		&	$0.45$	   & A.O.	 \\	
2. Almina ball    & 	$1.0\pm0.1$ &	3.9		&	$0.45$ 	& A.O.  \\	
3. Zirconia ball  &	$0.5\pm0.1$ &	5.9		&	$0.46$    & A.O.	 \\	
4. Rough sand	  &	$1.0\pm0.3$ &	2.6		&	$0.65$    & JIS	 \\	
\hline
\end{tabular}
\caption{Granular media used in the experiments. A.O. and JIS represent AS ONE (supplier of materials) and Japanese Industrial Standards.}
\label{tab:grain_property}
\end{center}
\end{table}

\begin{figure*}[htbp]
\begin{center}
\includegraphics[width=6.5 in]{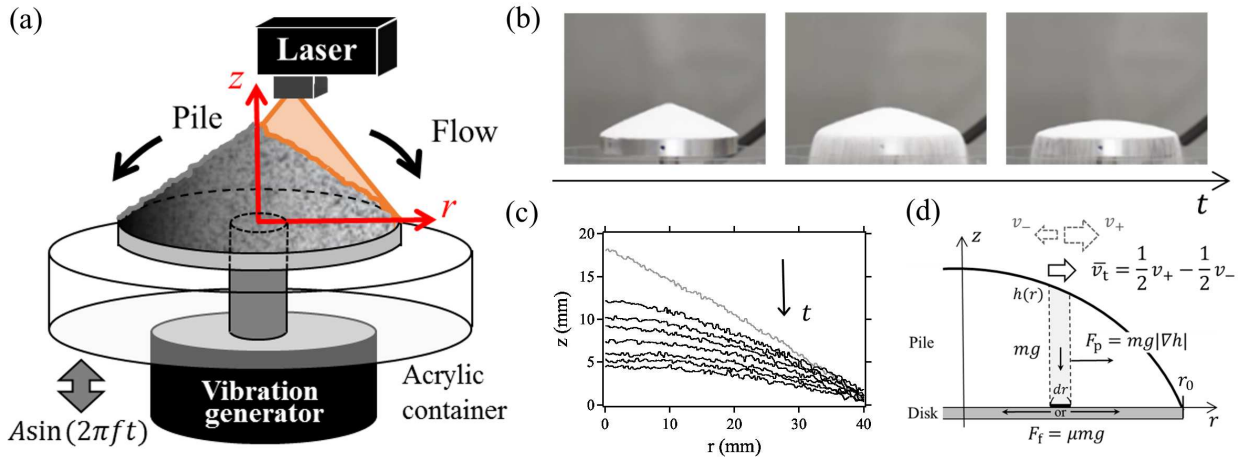}
\end{center}
\caption{(a) Schematic illustration of the experimental system.
(b) Snapshots of a relaxing pile during vibration.
(c) Raw data of radial profiles taken by a laser profiler. A gray curve shows a profile of an initial pile with the angle of repose, which is taken before turning on a vibration generator. Black profiles are taken at $t=0, 0.26, 0.5, 1.0, 2.0, 3.0, 5.0$~s from top to bottom. Note that the data between gray and top black profiles are not used in this study as the vibration amplitude $A$ is not constant.
(d) Image of the NDT model. In the main text, the forces applied to the filled thin slice are discussed.}
\label{fig:exp_system}
\end{figure*}

A schematic illustration of the experimental system is given in Fig.~\ref{fig:exp_system}(a).
A conical pile with the angle of repose is created on a disk using each set of granular materials with diameter $d=0.5~{\rm or}~1.0$~mm, density $\rho=2.6\sim5.9$~g/cm$^3$, and $\rm{tan}\theta_{\rm c}=0.45\sim0.65$ (Table~\ref{tab:grain_property}).
The experimental system is mounted on an electromechanical vibrator (EMIC 513-B/A).
The radius of the disk $r_0$ is $40$ mm, and a layer of the same grain is glued on its surface.
Then, sinusoidal vertical vibration is continuously applied to the disk.
The amplitude $A$ and frequency $f$ are varied in the range of $10^{-4}\sim10^{0}$~mm and $50\sim500$~Hz such that granular piles are destabilized.
Actually, it is difficult to precisely determine the onset criterion of the fluidization.
Although the maximum acceleration scaled by the gravitational acceleration $\Gamma=A(2\pi f)^2/g$ seems to be the most relevant, the critical value $\Gamma_{\rm c}$ fluctuates around $1<\Gamma_{\rm c}<2$ depending on $f$~\cite{Tennakoon1998, King2000}.
To avoid such complexity and focus only on clearly fluidized regimes, in this study a pile is subjected to relatively strong vibration of $2\leq\Gamma\leq10$.
We perform experiments for each set of conditions three times to check the reproducibility.
Once granular media begin to flow, the shape of the pile is relaxed (Fig.~\ref{fig:exp_system}(b)).
Outflowed grains are collected by an acrylic container surrounding the disk.
As a unique advantage of this setup, the sidewall effect, which is difficult to be removed in usual quasi-2D flows~\cite{Jop2005, Jop2006}, does not appear at all.
To measure the flow properties during the relaxation, surface profiles of the pile are continuously recorded by a high-speed laser profiler (KEYENCE LJ-V7080).
Figure~\ref{fig:exp_system}(c) shows profiles taken in the experimental condition of $A=0.04~{\rm mm}$ and $f=200~{\rm Hz}$ ($\Gamma=6$) with Material 1 in Table~\ref{tab:grain_property}.
The measurement ranges are from the center to the edge of the pile along the radial direction ($r=0\sim40~{\rm mm}$).
The origin of the height coordinate $z=0$ is calibrated to the surface of the disk.
In this experiment, the amplitude of vibration is gradually increased during the initial $0.5$~s to calmly get to stable vibration conditions without burst signals.
The time when the vibration achieves a stable state is defined as $t=0$~s.
Henceforth, these experimental conditions are used for subsequent plots unless otherwise noted.

\begin{figure}[t]
\begin{center}
\includegraphics[width=80mm]{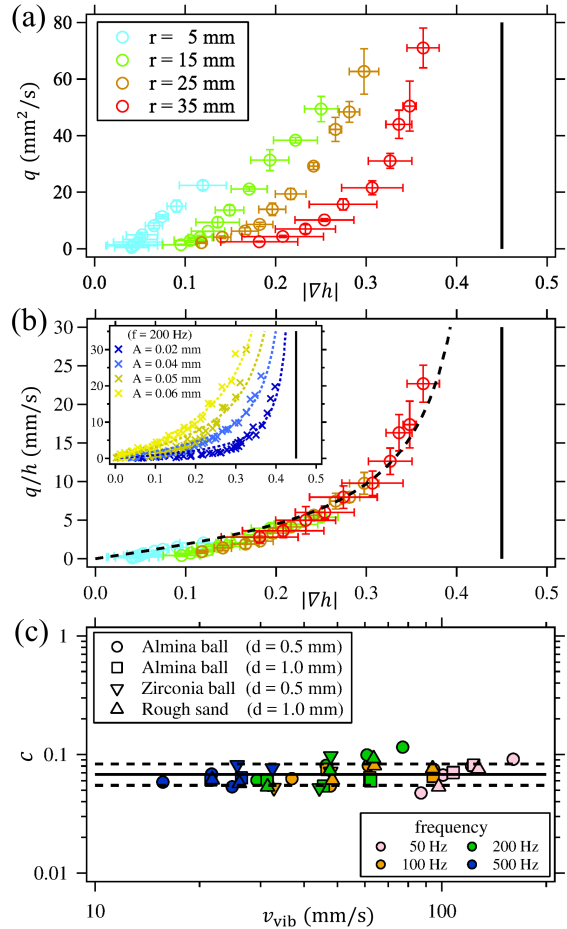}
\end{center}
\caption{
(a) Flux $q$ versus slope $|\nabla h|$.
The vertical solid line corresponds to $|\nabla h|={\rm tan}\theta_{\rm c}$.
Colors represent analysis points.
(b) Depth-averaged velocity $q/h(=\overline{v_{\rm t}})$ versus slope $|\nabla h|$.
Colors are identical to those in (a).
A dashed curve shows the best fitting by Eq.~(\ref{eq:NDT_model}).
Note that $\mu$ is fixed at tan$\theta_{\rm c}$.
Inset: Analysis results for various vibration conditions.
The axises are the same as the main plot.
All the data are also fitted by Eq.~(\ref{eq:NDT_model}).
(c) Relaxation efficiency $c$ as a function of the maximum vibration velocity $v_{\rm vib}$.
Colors and symbols represent vibration frequency $f$ and materials used (Table.~\ref{tab:grain_property}).
Solid and dashed lines depict $c=0.068$ with $1\sigma=0.014$.
}
\label{fig:result}
\end{figure}

\textcolor{red}{In the analysis, to investigate the flow property, the flux and slope are measured at four points with interval $\Delta r=10$~mm ($r=5, 15, 25, 35$~mm) for a variety of time.
Since the pile is relaxed axisymmetrically, the flux $q$ at $r=r'$ and $t=t'$ is calculated as
\begin{equation}
q(r',t')=\frac{1}{r' dt} \int_0^{r'} |h(r, t'+dt)-h(r, t')|r dr,
\label{eq:flux_definition}
\end{equation}
where $h(r, t)$ is the height of the pile at position $r$ and time $t$, which is measured from $z=0$~\textcolor{red}{\cite{SM}}}.
Since the relaxation dynamics slows down as time goes on, logarithmically-increasing time bins are employed for $dt$\textcolor{red}{~\cite{SM}}, i.e., $dt=0.1\times{\sqrt{2}}^n~\rm{s}~(n=0, 1, 2, \cdots)$.
\textcolor{red}{The slope $|\nabla h|(=|\partial h/\partial r|)$ at position $r$ is calculated by the linear least-squares method using profiles from $r - \Delta r/2$ to $r + \Delta r/2$.}
The relation between $q$ and $|\nabla h|$ is shown in Fig.~\ref{fig:result}(a).
\textcolor{red}{Following the simplest assumption, Fick's law of diffusion, $q$ is proportional to $|\nabla h|$. 
The data trend, however, suggests nonlinearity and that $q$ is not scaled only by $|\nabla h|$ but also depends on $r$.}
To explain this complex dependence and reproduce the granular-heap relaxation, the transport model will be derived below.

First of all, let us consider the dynamics of a thin annular slice of granular media with width $dr$ and height $h(r)$ (Fig.~\ref{fig:exp_system}(d)).
\textcolor{red}{For the sake of simplicity, we assume that the pile is uniformly fluidized throughout the layer due to relatively strong vibration with $\Gamma > 1$; 
and the slice is horizontally transported on a vibrating plate like a solid block~(cf.~\cite{Sanchez2007}).
This idea is contrastive to the shear-band structure assumed by Roering {\it et al}.~\cite{Roering1999}, i.e., the fluidized layer is localized only around the surface.}
The forces applied to a slice in motion consist of two factors: the force due to the pressure gradient $F_{\rm p}=mg|\nabla h|$ and the frictional force between the slice and disk $F_{\rm f}=\mu mg$, where $m=\rho_{\phi} h(r) 2\pi r dr$, $\rho_{\phi}$ is the bulk density of granular media, and $\mu$ is a coefficient of friction.
The former always works along the positive direction of the $r$-axis, while the latter is applied along the opposite direction of the motion.
\textcolor{red}{Note that hydrostatic pressure is assumed as the pile is completely fluidized; the effects of air drag and cohesion are neglected as constitutive grains are sufficiently large, dense, and dry.}

As a next step, based on the idea proposed by Roering {\it et al}.~\cite{Roering1999}, we assume that the oscillating disk supplies energy $\Delta E$ to the pile during infinitesimal time $\Delta t$ isotropically such that the slice is able to move along both directions with equal probabilities.
In other words, during $\Delta t$ the vibration attempts to transport the slice to the positive direction of the $r$-axis with velocity $v_{+}$ and probability $1/2$, while to the negative direction with velocity $v_{-}$ and probability $1/2$ as well.
This process continues as long as vibration is applied.
The similar modeling has also been developed for coarsening dynamics in a vibro-fluidized compartmentalized granular gas~\cite{Meer2004}.
The important point in this modeling is that $v_{+}$ is larger than $v_{-}$ because the resistant force against the motion to the positive direction $F_{+}=F_{f}-F_{p}$ is smaller than that against the negative direction $F_{-}=F_{f}+F_{p}$.
As a result, the average transport velocity along the $r$-axis $\overline{v_{\rm t}}$ can be written as
\begin{eqnarray}
\overline{v_{\rm t}} &=& \frac{1}{2}v_{+}-\frac{1}{2}v_{-}   \nonumber\\
&=& \frac{1}{2}\frac{\Delta E}{\Delta t} \left(\frac{1}{F_{+}} - \frac{1}{F_{-}}\right) \nonumber\\
&=& \frac{W}{2mg} \left(\frac{1}{\mu-|\nabla h|} - \frac{1}{\mu+|\nabla h|}\right),
\label{eq:transport_velocity}
\end{eqnarray}
where $W=\Delta E/\Delta t$ means the average work done per unit time.
Here, let us suppose that a certain percentage of the vibration energy inputted vertically is used for the horizontal transport of the slice.
This idea allows $W$ to take the following simple form:
\begin{eqnarray}
W=c m g v_{\rm vib},
\label{eq:work_rate}
\end{eqnarray}
where $c$ is a constant, and $v_{\rm vib}=A(2\pi f)$ represents a characteristic vibration velocity.
Using the above expressions, $\overline{v_{\rm t}}$ can finally be rewritten as
\begin{equation}
\overline{v_{\rm t}} = \frac{cv_{\rm vib}}{\mu^2}\frac{|\nabla h|}{1-(|\nabla h|/\mu)^2}.
\label{eq:NDT_model}
\end{equation}
In this model, when $|\nabla h| \ll \mu$, $q(=h \overline{v_{\rm t}})$ is almost proportional to $|\nabla h|$, which can be interpreted as Fick's law of diffusion.
On the other hand, $q$ diverges rapidly as $|\nabla h|$ approaches $\mu$, indicating strong nonlinearity.
In the following, we call Eq.~(\ref{eq:NDT_model}) the Nonlinear Diffusion Transport (NDT) model.

Let us verify whether the NDT model can consistently explain the experimental data.
According to the above modeling, $q/h$ corresponds to $\overline{v_{\rm t}}$ which obeys Eq.~(\ref{eq:NDT_model}).
If the NDT model is correct, $\overline{v_{\rm t}}=q/h$ must be a function of only $|\nabla h|$ in contrast to $q$ which depends on both $|\nabla h|$ and $r$ (Fig.~\ref{fig:result}(a)).
Figure~\ref{fig:result}(b) shows the relation between $q/h$ and $|\nabla h|$.
As expected, all the data are collapsed into a single curve, which can be fitted by Eq.~(\ref{eq:NDT_model}), where the fitting parameter is only $c$.
$\mu$ is fixed at tan$\theta_{\rm c}$ in Table~\ref{tab:grain_property}.
In spite of this simplification, the scaling of Fig.~\ref{fig:result}(b) is universal to all experimental data.
The evidence is given in the inset of Fig.~\ref{fig:result}(b), which shows analysis results for various vibration conditions.
All the data are also fitted well by Eq.~(\ref{eq:NDT_model}).

Another point that needs to be confirmed is the parameter dependence of $c$.
The values of $c$, which are calculated by the least-squares fitting to Eq.~(\ref{eq:NDT_model}), are plotted as a function of $v_{\rm vib}$ for four granular materials (Table~\ref{tab:grain_property}) in Fig.~\ref{fig:result}(c).
$c$ seems to be independent of any experimental conditions and shows the constant value $0.068\pm0.014$.
Since $c$ indicates a velocity conversion rate from the input vertical vibration to the horizontal transport, it is natural that $c$ is less than $1$ and uniquely determined by the system itself.
However, one can speculate that $c$ might depend on the mechanical properties of constituent grains (e.g., softness) and/or boundary conditions of the system. 
Additional studies to evaluate these effects are important future works.

\begin{figure}[t]
\begin{center}
\includegraphics[width=80mm]{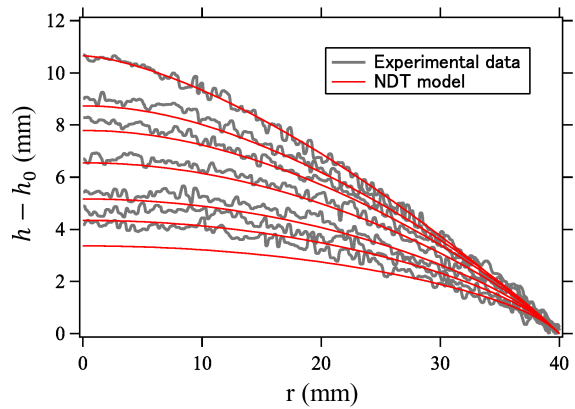}
\end{center}
\caption{Comparison between profiles taken by a laser profiler and those computed by the NDT model.
The initial shape for the model computation is approximated by \textcolor{red}{$h-h_0\sim({r_0}^{1.6}-r^{1.6})$} with $r_0=40$~mm.
\textcolor{red}{The time step and spatial resolution of the numerical calculation are $10^{-5}$ s and $0.5$ mm, respectively.} 
Experimental and computed curves agree well in relatively steep regimes.
When the shape of the pile get flattened to some extent, granular media reach a jammed state, where the relaxation almost halts.}
\label{fig:profile}
\end{figure}

Next, in order to reproduce the relaxation process of the pile, we numerically solve the equation of continuity with $q=h\overline{v_{\rm t}}$ given by the NDT model:
\begin{equation}
\frac{\partial h}{\partial t} = \nabla \cdot \left(h \frac{cv_{\rm vib}}{\mu^2}\frac{\nabla h}{1-(|\nabla h|/\mu)^2}\right).
\label{eq:NDT_model_num}
\end{equation}
\textcolor{red}{Note that the following boundary conditions are imposed from the experimental configuration: $q=0~{\rm mm}^2/$s at $r=0~{\rm mm}$ and $h_0(t)=h(t,r_0)$, where $h_0(t)$ is the thickness of a granular layer flowing out of a disk at a given time, which typically consists of a few grain diameters.} 
The comparisons between profiles taken by a laser profiler and those computed by the NDT model are shown in Fig.~\ref{fig:profile}, which are in good agreement with each other.
These facts support the validity of the NDT model.
Interestingly, Eq.~(\ref{eq:NDT_model_num}) does not include the dependence on grain size $d$.
This is because the relevant length scale governing the relaxation dynamics is the height $h(r)$ rather than $d$ in a fully-fluidized pile.

Here, limitations of the NDT model are briefly discussed.
\textcolor{red}{First, according to the rheological models for dense fluidized flows, granular temperature could also affects the transport velocity (e.g.,~\cite{Jenkins2012}). 
Although this effect is expected to be dominantly linked to vibration velocity, the correspondence has not yet been reached.
The detail comparison with such rheological models is necessary to improve the model.}
Second, we observed that the pile reaches a ``jammed'' state leaving a finite slope even if submitted to vibration for a long time.
As can be seen in both Figs.~\ref{fig:result}(b) and \ref{fig:profile}, model curves exhibit misfits with experimental data when slopes approach zero.
These jammed states become more apparent as $v_{\rm vib}$ decreases.
In this regime, inertial energy supplied to grains is insufficient to overcome potential barriers of their neighbors~\cite{Jaeger1989, Roering2004}.
Conversely, when the vibration is too strong, the transition into a granular-gas phase~\cite{Jaeger1996} must occur, in which the NDT model is no longer suitable.
The best we can say at present is that the NDT model can be applied only in the restricted range (at least $v_{\rm vib}=10\sim200~{\rm mm/s}$) except for low-angle conditions.

\textcolor{red}{
Finally we show a simple application of the NDT model toward the crater relaxation process caused by impact-induced seismic shaking (Fig.~\ref{fig:application}).
Let us consider the crater erasure by a meteorite impact with diameter $4$~m onto asteroid Itokawa.
By this impact, the entire surface of Itokawa is shaken with $\Gamma>1$~\cite{Yamada2016}.
The expected typical vibration velocity and duration are $V_{\rm seis} \sim 10^{-2}~{\rm m/s}$ and $T_{\rm seis} \sim 10^1$~s~\cite{Yamada2016,SM}. 
Then, the crater of diameter $D_{\rm cra}$ and depth $H_{\rm cra}$ is erased when the mass flow exceeds the crater volume. 
The order estimate of this mass balance is expressed as $\overline{v_{\rm t}}H_{\rm cra} D_{\rm cra} T_{\rm seis} \sim D_{\rm cra}^2 H_{\rm cra}$. 
By neglecting the nonlinear term in Eq.~(\ref{eq:NDT_model}) which is important only for steep slopes, $D_{\rm cra} \sim cV_{\rm seis}\mu^{-2} |\nabla H| T_{\rm seis}$ holds, where the typical $\mu$ value is tan$35^\circ$~\cite{Fujiwara2006} and $|\nabla H| \sim H/D \sim 10^{-1}$~\cite{Hirata2009,Katsuragi2016}. 
Substituting specific values into this relation, the maximum crater diameter erased by this impact-induced shaking is estimated as $D_{\rm cra}\sim 10^{-3}$~m. This value is comparable with actual grain (regolith) size~\cite{Fujiwara2006}, which means that the the accumulation of multiple impacts is necessary to erase large craters. 
To estimate the relaxation time scale more precisely, much more careful computation as discussed in Ref.~\cite{Yamada2016} must be performed. 
}

In summary, we have performed experiments to understand the relaxation dynamics of granular-heap structure on a vertically-vibrated plate.
To explain the experimental results, we have proposed the Nonlinear Diffusion Transport (NDT) model, which describes the flux of granular particles including the dependence on experimental parameters.
The fitting parameter is only the relaxation efficiency $c$, which does not depend on experimental conditions.
\textcolor{red}{Considering this universality, from a rheological point of view, our experimental setup could provide a new insight to characterize frictional properties of flowing granular media under vibration.}
We also believe that the framework of our modeling for relaxation is potentially applicable to other experimental configurations with different heap sizes or perturbation types.

\begin{acknowledgments}
We would like to thank S. Watanabe, H. Kumagai, S. Sirono, and T. Morota for fruitful discussions.
This study has been supported by JSPS KAKENHI No.~15H03707, No.~16H04025, No.~17H05420, and No.~17J05552.
\end{acknowledgments}

\bibliography{sandpile}

\end{document}